\documentclass{optica-article}

\journal{optcon}

\articletype{Research Article}

\usepackage{mathrsfs}
\usepackage{graphicx,setspace,lscape,longtable,xr}
\usepackage[square,sort,comma,numbers]{natbib}
\usepackage{mathrsfs,amsmath,color}

\usepackage{amssymb}

\usepackage{amsthm}
\usepackage{epsfig,graphicx,pdfpages}
\usepackage{rotating}
\usepackage{float}
\usepackage{bm}
\usepackage{CJK}
\usepackage{ragged2e}
\usepackage{setspace}
\usepackage{threeparttable}
\usepackage{multirow}
\usepackage{url}
\usepackage{caption}
\usepackage[labelformat=simple]{subcaption}

\usepackage{multirow}

\usepackage{lineno}

\begin{document}

\title{High-Resolution Boundary Detection for Medical Image Segmentation with Piece-Wise Two-Sample T-Test Augmented Loss}

\author{Yucong Lin\authormark{1*}, Jinhua Su\authormark{5*}, Yuhang Li\authormark{3}, Yuhao Wei\authormark{3}, Hanchao Yan\authormark{5}, Saining Zhang\authormark{3}, Jiaan Luo\authormark{3}, Danni Ai\authormark{4}, Hong Song\authormark{3}, Jingfan Fan\authormark{4}, Tianyu Fu\authormark{1}, Deqiang Xiao\authormark{4}, Feifei Wang\authormark{2,5**}, Jue Hou\authormark{6**}, Jian Yang\authormark{4**}}

\address{
\authormark{1}School of Medical Technology, Beijing Institute of Technology, Beijing, 100081, China\\
\authormark{2}Center for Applied Statistics, Renmin University of China, Beijing, 100872, China\\
\authormark{3}School of Computer Science and Technology, Beijing Institute of Technology, Beijing, 100081, China\\
\authormark{4}Beijing Engineering Research Center of Mixed Reality and Advanced Display, School of Optics and Photonics, Beijing Institute of Technology, Beijing, 100081, China\\
\authormark{5}School of Statistics, Renmin University of China, Beijing, 100872, China\\
\authormark{6}School of Public Health, University of Minnesota, Minneapolis, MN 55455, USA\\
\authormark{*}The authors are equally contributed\\
\email{\authormark{**}Corresponding authors: feifei.wang@ruc.edu.cn, hou00123@umn.edu, jyangbit@163.com}}

\begin{abstract}
Deep learning methods have contributed substantially to the rapid advancement of medical image segmentation, the quality of which relies on the suitable design of loss functions. Popular loss functions, including the cross-entropy and dice losses, often fall short of boundary detection, thereby limiting high-resolution downstream applications such as automated diagnoses and procedures. We developed a novel loss function that is tailored to reflect the boundary information to enhance the boundary detection. As the contrast between segmentation and background regions along the classification boundary naturally induces heterogeneity over the pixels, we propose the piece-wise two-sample t-test augmented (PTA) loss that is infused with the statistical test for such heterogeneity. We demonstrate the improved boundary detection power of the PTA loss compared to benchmark losses without a t-test component.
\end{abstract}

\section{Introduction}

Medical image segmentation plays an important role in biomedical research because it often serves as the primary task in many medical applications \citep{Murugesan2019Psinet}. Fully automatic segmentation methods, such as liver and liver tumor segmentation, brain and brain tumor segmentation, optic disc segmentation, cell segmentation, lung segmentation, pulmonary nodule segmentation, and cardiac image segmentation \citep{wang2022medical}, are essential for the diagnosis of serious diseases \citep{Chen_2019_CVPR}. Therefore, it is important to improve the efficiency and accuracy of medical image segmentation methods.

Medical image segmentation involves segmenting specific organs (e.g., the pancreas, liver, and bladder), determining certain functional parts of an organ (e.g., cardiac segmentation and retinal vessel segmentation), and identifying tumors in the organs. Medical images can generally be categorized according to the imaging technology and data form. Imaging technology includes X-ray photos, computed tomography, magnetic resonance imaging (MRI), and ultrasound imaging. Raw measurements are transformed into pixelated imaging data as part of the standard process. Although the original data are mostly three-dimensional images, two-dimensional slices are often created according to clinical procedure protocols that target specific applications. Most medical image segmentation methods are designed for two-dimensional slices.

Recent medical image segmentation methods are increasingly based on deep learning owing to its outstanding performance, among which convolutional neural networks (CNNs) has a substantial share. The seminal U-Net \citep{Ronneberger2015} is a U-shaped CNN. Low-level features (representing the localization features) and high-level features (representing the context features) can be integrated using the concatenation between encoder and decoder layers of the same depth. Moreover, U-Net has a shallow architecture with fewer parameters to estimate, which makes it particularly suitable for small medical image datasets \citep{Chen_2019_CVPR}. The loss function is another key perspective for improving the effectiveness of deep learning methods. The cross-entropy loss, weighted cross-entropy loss, and dice loss (also known the dice similarity coefficient) are widely used loss functions in image segmentation tasks \citep{wang2022medical}. All these loss functions focus on the pixels of the entire patch instead of the boundary region, which means that they are region based and lack boundary (contour or shape) information. However, the boundaries of objects in medical image segmentation may be fuzzy or noisy. As a result, although these loss functions can provide a roughly correct region, the boundaries may be twisted when noise is encountered.

In recent years, an increasing number of studies have focused on the boundary regions \citep{Chen_2019_CVPR, Le2021OffsetCurve}. The heterogeneity of the pixels in the boundary region can provide useful information for image segmentation. In fact, traditional segmentation methods of differential operators that identify the edge according to its sharp rate of changes in color offer another perspective to use the heterogeneity over the pixels in the boundary region, but these methods also have limitations. For example, the derivatives are highly sensitive to local noise and are not ideal for discrete pixels because they are defined as continuous objects. To overcome these challenges, various studies have been conducted to improve the performance of medical image segmentation by making full use of the boundary information.

One means of improving the performance of medical image segmentation in the boundary region is to modify the network architecture. Research in this regard can generally be divided into three categories. The first category is weighting the boundary region to reinforce the boundary representation of the features. In this category, the attention mechanism is a type of weighting method that focuses on ambiguous boundary regions \citep{Fan2020PraNet,Fan2020InfNet,Wang2021boundaryawaretransformer}. Moreover, methods are available for learning a map of confidence or residual representation to change the weights of the pixels \citep{Wang2019BoundaryAdversarial,Wang2021residual,Zhang2021}.
The second category involves designing specialized encoding modules to extract the boundary features, such as selective multi-kernel encoder methods and parallel encoder competitor solutions \citep{Fang2019SelectiveFeature,Murugesan2019Psinet,Yu2019EAGAN,Wang2021BoundaryCoding}, encoders with deep supervision methods \citep{Yang2019RNNdeepsupervision,Gu2021kCBAC}, skip connections of concatenations, or residual convolution paths \citep{Fang2019SelectiveFeature,Seo2020mUnet,Wang2021boundaryawareUnet}. The third category involves adapting deep learning networks to fit traditional geometric feature methods, such as the Canny edge fusion method \citep{Wang2021boundaryawareUnet}, active contour method \citep{Ma2021Geodesic}, and oriented derivative method \citep{Cheng2021OrientedDerivative}.

Another research direction focuses on modifying the loss functions. Studies in this area have proposed additional fitting terms for the loss function. Some of the added terms are constraint derived from distance measure, like the Hausdorff distance. For example, Marc-Adrien et al.\cite{MarcAdrien2021Bladderloss} applied the average surface distance and Hausdorff method by adapting them into differentiable forms and adding them to the traditional loss function. Other notable studies include Chen et al.\cite{Chen_2019_CVPR}; Kervadec et al.\cite{pmlr-v102-kervadec19a}; Zheng et al.\cite{Zheng2020deeplevelset} and Le et al.\cite{Le2021OffsetCurve}. Inspired by the active contour model \citep{ACMChan1999}, the authors of these works rewrote the energy function to create a new constraint. This constraint contains the square term of the pixel-to-boundary curve inside and outside the boundary, which can reflect the inside and outside differences of the boundary and the difference between the computed and ideal boundaries.

Based on the statistical hypothesis testing concept, we propose a new loss function that exploits the heterogeneity across boundaries. As boundary indicates the separation of distinct objects, the gray values inside and outside the boundary should differ. We adopt the two-sample t-test to detect this difference effectively. The two-sample t-test is a classical statistical hypothesis testing method that can determine whether two observation groups are significantly different. In this study, we incorporate the two-sample t-test into the loss to contrast the gray levels on opposite sides of the estimated boundary. We divide the entire boundary into local pieces to adapt to changes in patterns along both sides of the boundary, and subsequently, conduct separate two-sample t-tests within the neighborhood of each piece, which results in the proposed piecewise two-sample t-test augmented (PTA) loss. We consider the infusion of the PTA into several existing loss functions (e.g., the cross-entropy and dice loss) to optimize various medical image segmentation models.

The remainder of this paper is organized as follows: The PTA loss is introduced in detail in Section 2. Section 3 presents a series of simulation studies to demonstrate the correct action direction of the PTA loss. Section 4 outlines the application of the PTA loss, along with other traditional loss functions, to a public cardiac cine MRI dataset to reveal the superior detection performance of the PTA. Further discussions and concluding remarks are provided in Section 5.

\section{PTA LOSS}

\subsection{Problem Formulation}
\label{subsection:formulation}
Medical image segmentation usually involves two basic tasks: the detection of specific organs and the subsequent accurate segmentation thereof from the image. In medical image segmentation, these two tasks are generally unified by one task of \emph{pixel-wise classification}, to be trained with the initial image and its corresponding ground-truth mask. Putting the classification model on a testing image, we will generate the prediction on whether each pixel in the image is in the \emph{segmentation region} or \emph{background region}.

\begin{figure}[h]
    \centering
    \begin{subfigure}[t]{.3\textwidth}
        \centering
        \includegraphics[width=\textwidth]{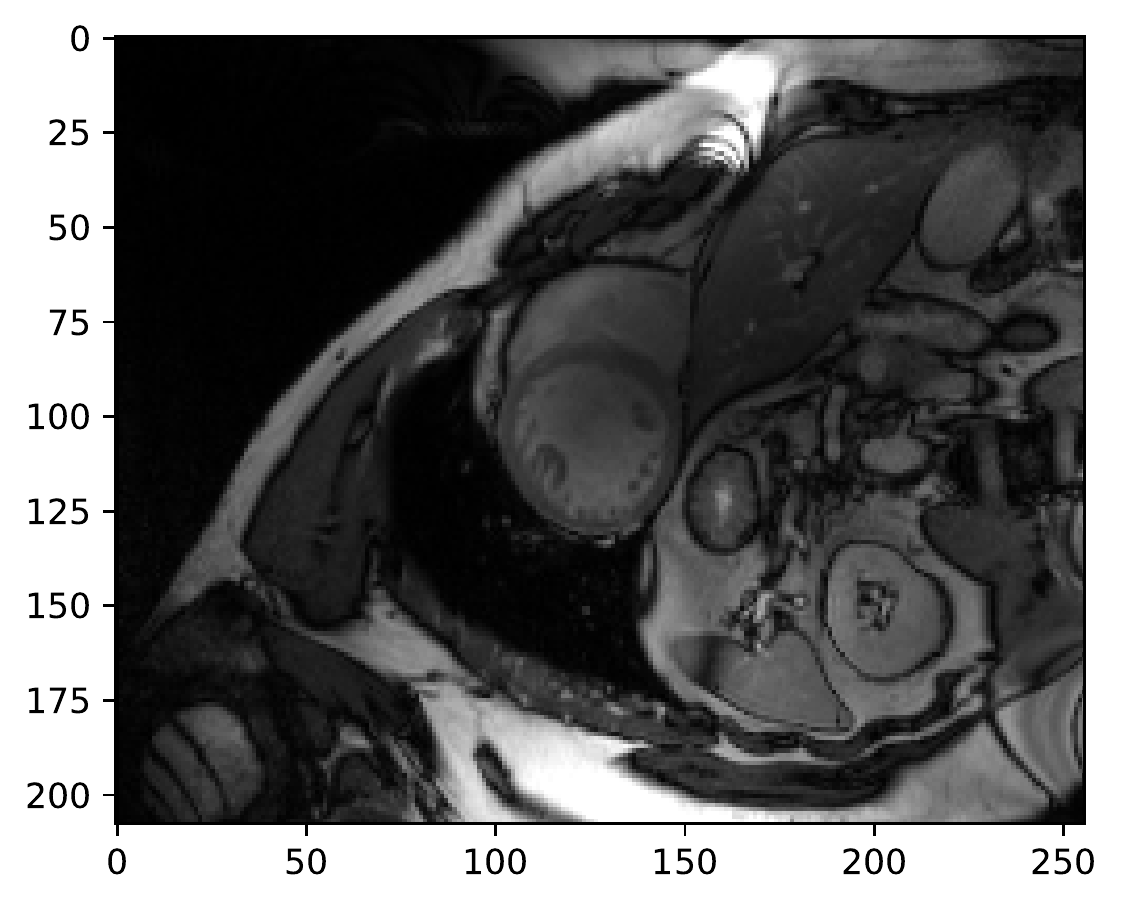}
        \caption{Initial image}
        \label{fig:initial-image}
    \end{subfigure}
    \quad
    \begin{subfigure}[t]{.3\textwidth}
        \centering
        \includegraphics[width=\textwidth]{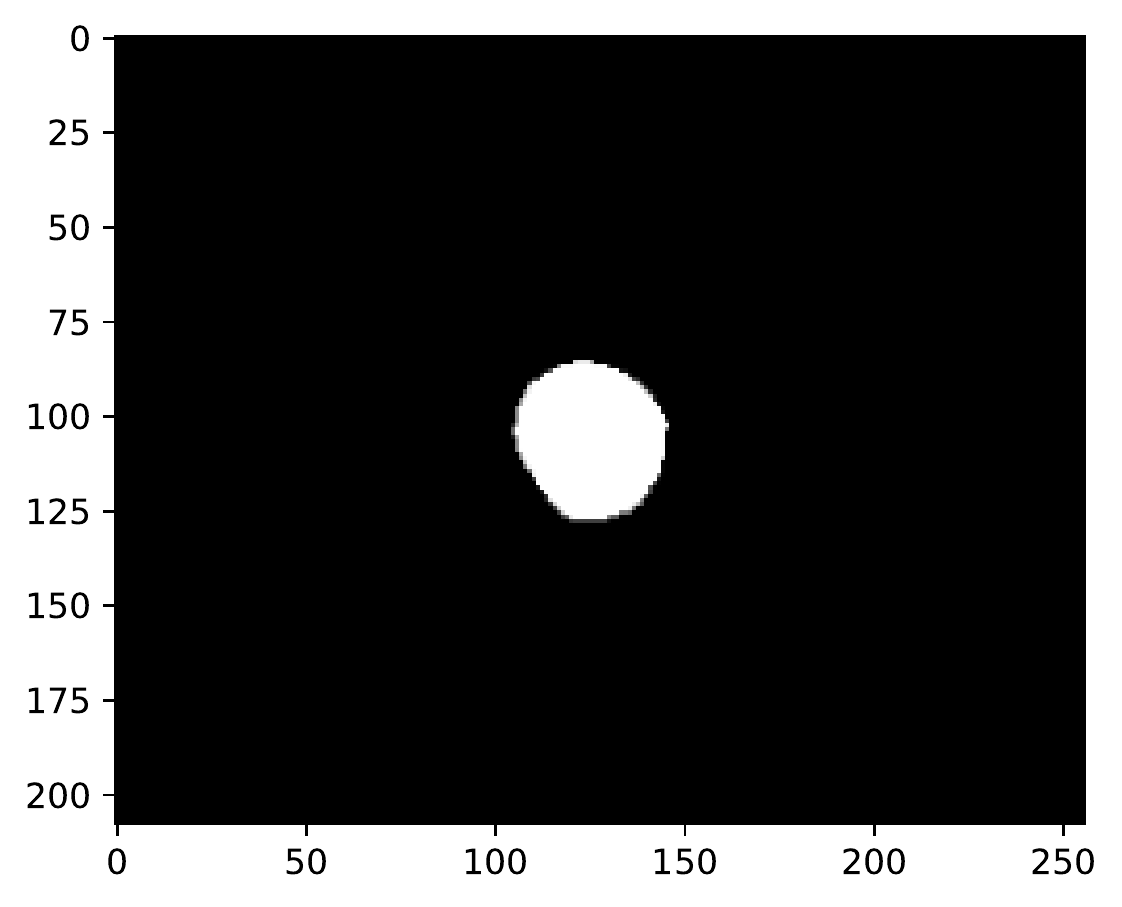}
        \caption{Ground truth}
        \label{fig:ground-truth}
    \end{subfigure}
    \quad
    \begin{subfigure}[t]{.3\textwidth}
        \centering
        \includegraphics[width=\textwidth]{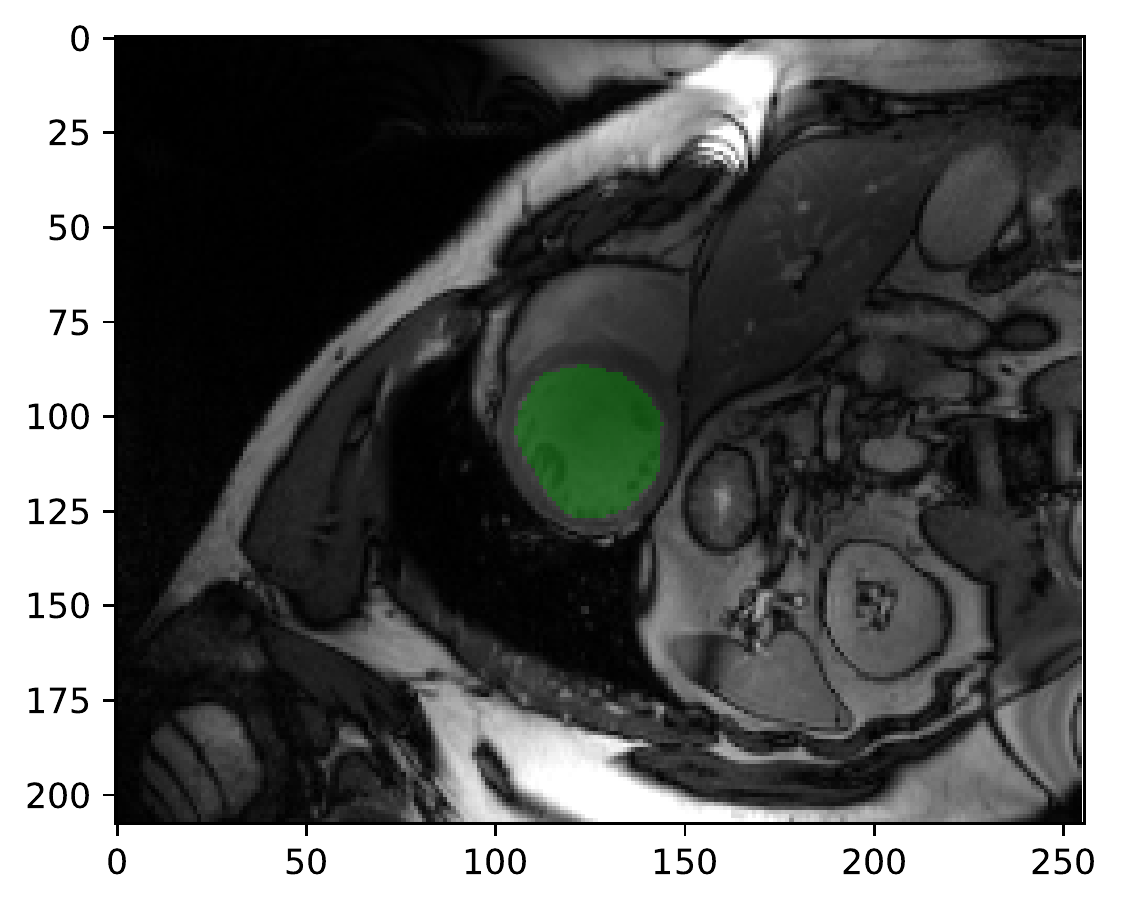}
        \caption{Image with its ground-truth region}
        \label{fig:image-and-ground-truth}
    \end{subfigure}
    \caption{Image example and its ground-truth region from cine MRI data.}
    \label{fig:initial-image-and-gt}
\end{figure}

In the following, we define several notations and provide a clear definition of the segmentation problem. We assume that there exists a spatial domain $\Omega$ that contains a set of pixel positions. Let $\boldsymbol{x}\in\Omega$ denote the position of any arbitrary pixel that belongs to $\Omega$. Subsequently, an image can be represented as a set of pixels for which the color values are defined on $\Omega$. In this study, we focus on grayscale images, which can easily be converted from color images. In a grayscale image, each pixel is associated with a grayscale value $I(\boldsymbol{x})\in\mathbb{R}$. Subsequently, the image is defined as $I:\Omega\to\mathbb{R}$. The ground-truth masks, which indicate whether a pixel belongs to the background or segmentation region, should be provided in the pixel-wise classification task. We define the ground-truth label at each pixel as $g:\Omega\to\{0,1\}$ and the corresponding ground-truth region as $G=\{\boldsymbol{x}\in\Omega: g(\boldsymbol{x})=1\}$. In this task, a model is required to predict the probability that each pixel belongs to the segmentation region, which can be defined as $f:\Omega\to [0,1]$. Further, we use $S=\{\boldsymbol{x}\in\Omega: f(\boldsymbol{x})>\delta\}$ to denote the dichotomized segmentation region with a certain probability threshold $\delta$. For illustration purposes, Figure \ref{fig:initial-image-and-gt} presents an image example and its ground-truth region from the cine MRI data of Section 4.

The candidate segmentation should be measured using a loss function to train the classification model. Various loss functions have already been used in medical image segmentation tasks \citep{ma2021loss}. Classical loss functions include the cross-entropy loss (CE), weighted cross-entropy loss (WCE), and dice similarity coefficient loss (DSC). The detailed definitions of each loss function are as follows:
\begin{equation}
\begin{aligned}
    \mathcal{L}_{CE} = -\sum_{\boldsymbol{x}\in\Omega} g(x)\log f(x),\\
\mathcal{L}_{WCE}= -\sum_{\boldsymbol{x}\in\Omega}\omega(x)g(x)\log f(x),\\
\mathcal{L}_{DSC} = 1-2\frac{|G\cap S|}{|G|+|S|}.
\end{aligned}
\end{equation}

Optimal segmentation can be achieved by minimizing the above-mentioned loss functions. However, these global losses may sacrifice the accuracy at the boundary for clear interior and exterior pixels that dominate the image. Classical loss functions can reach its minimum when most pixels have been correctly classified. In this case, the refinement of the classification results at the boundary is poorly rewarded in the loss, whereas the collective impact on other pixels could adversely inflate the loss. Thus, boundary segmentation is often compromised with these losses. However, boundary pixels play an important role in the potential downstream tasks like automated measurement or excision. The segmentation accuracy of the target region should be acertained when attention is given the boundary region, mimicking the typical strategy of human radiologists. Therefore, the amplification of boundary information in loss functions is clearly a logical improvement for segmentation tasks.

\subsection{Two-Sample T-Test Loss for Boundary}

\begin{figure}[h]
    \centering
    \includegraphics[scale=0.55]{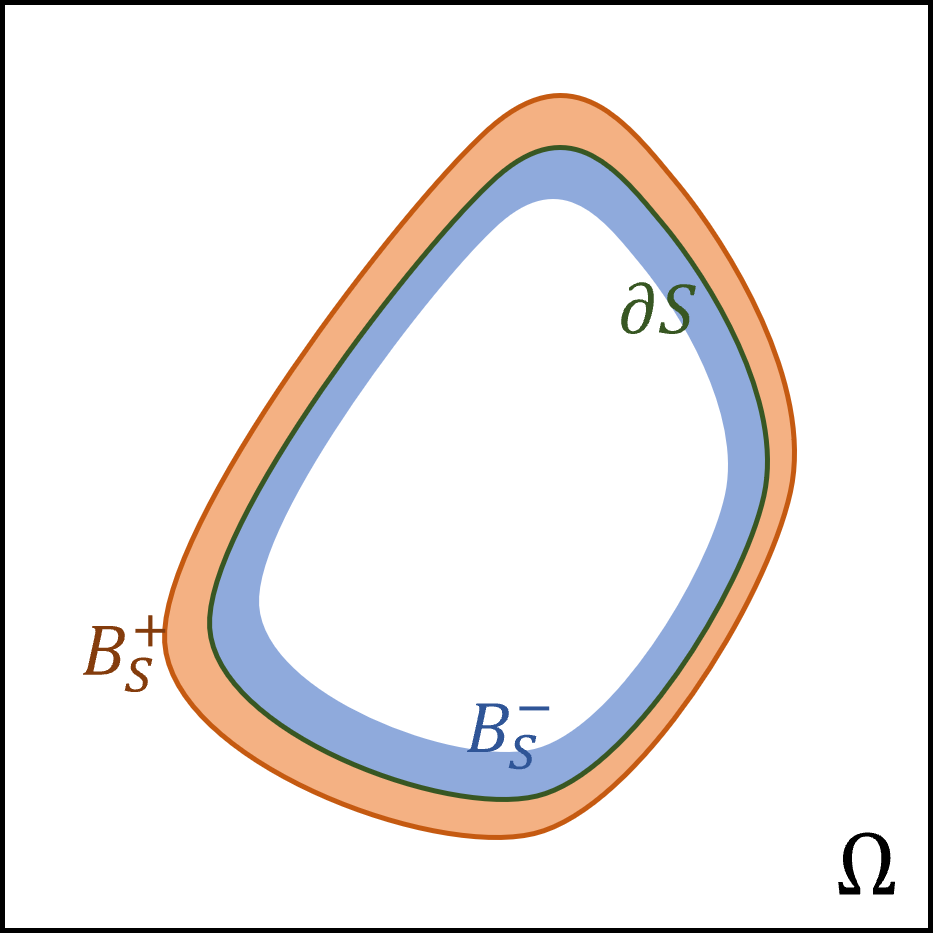}
    \caption{Illustrative example of $B_S^+$ and $B_S^-$ regions inside and outside of boundary $\partial S$.}
    \label{fig:b+b-_region}
\end{figure}

For a well posed segmentation problem, a significant contrast in grayscale values should exist between the \emph{inside} and \emph{outside} of the boundary that separates the segmentation and background regions. Motivated by this phenomenon, the boundary of the target region can be located at a curve that maximizes the contrast in grayscale values between the inside and outside of the boundary. Our PTA loss function is constructed based on this concept. First, given an estimated boundary from the classification model, we define two small regions around the boundary. As illustrated in Figure \ref{fig:b+b-_region}, we use $B^+_S$ and $B^-_S$ to indicate the adjacent \emph{outer} and \emph{inner} regions for the predicted segmentation region $S$ and its boundary $\partial S$, respectively. Choosing a buffer size $d$ always smaller than typical
(connected) segmentation region radius, we define
\begin{equation}
	B^+_S=\{\boldsymbol{x}\in \Omega\backslash S: d(\boldsymbol{x}, \partial S) < d\}, B^-_S=\{\boldsymbol{x}\in S: d(\boldsymbol{x}, \partial S) < d\}.
\end{equation}
\noindent
In the above, $d(\boldsymbol{x}, \partial S)$ indicates the distance from pixel $\boldsymbol{x}\in \Omega$ to boundary $\partial S$, and the distance refers to the Euclidean distance: $d(\boldsymbol{x},\partial S) = \min_{\boldsymbol{y}\in \partial S} ||\boldsymbol{x}-\boldsymbol{y}||_2$.


\begin{figure}[h]
    \centering
    \begin{subfigure}{.43\textwidth}
        \centering
        \includegraphics[width=\textwidth]{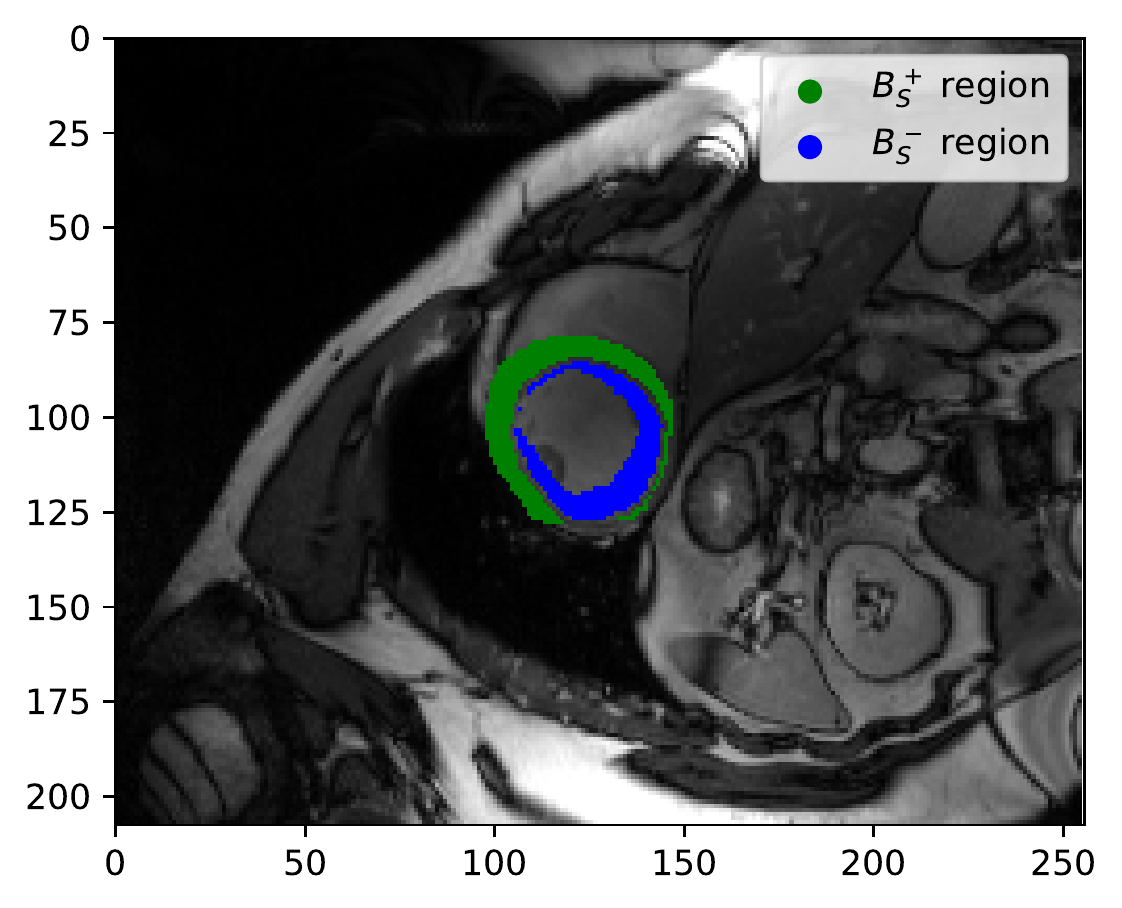}
        \caption{Example of $B^+_S$ and $B_S^-$}
        \label{fig:B+B-visual}
    \end{subfigure}
    \quad
    \begin{subfigure}{.53\textwidth}
        \centering
        \includegraphics[width=\textwidth]{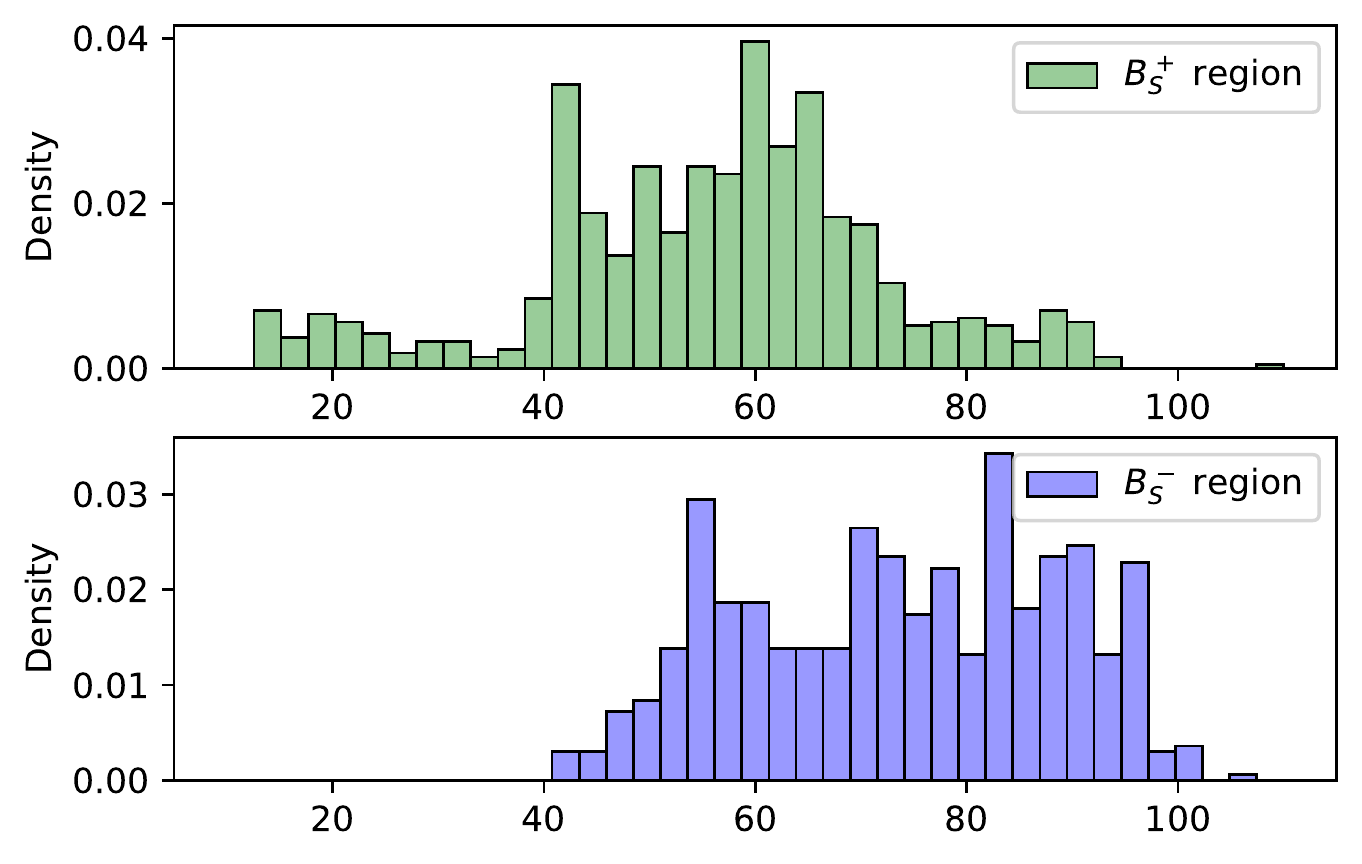}
        \caption{Grayscale value distributions in $B^+_S$ and $B_S^-$}
        \label{fig:B+B-pixel-dist}
    \end{subfigure}
    \caption{Visualization of $B_S^+$ and $B_S^-$ regions and grayscale value distributions in two regions}
    \label{fig:B+B-visual-and-comp}	
\end{figure}

An example of the $B_S^+$ and $B_S^-$ regions in a medical image is depicted in Figure \ref{fig:B+B-visual}. For further verification of the idea that a difference exists between the distributions of the grayscale values of the pixels \emph{inside} and \emph{outside} the boundary, Figure \ref{fig:B+B-pixel-dist} presents the two distributions using histograms. The figure indicates clear differences in the distributions of the pixels in the $B_S^+$ and $B_S^-$ regions. This finding demonstrates that boundary information can be used to improve the segmentation in the model.


A natural approach for measuring the difference between the distributions of the grayscale values in $B_S^+$ and $B_S^-$ is to compare the means of the grayscale values in the two regions directly. Specifically, we define the difference in the grayscale values in the two regions as follows:
\begin{equation}
\label{eq:spta-idea}
    v^*=|\bar{v}_+ -\bar{v}_-|\text{,}
\end{equation}
where $\bar{v}_+$ and $\bar{v}_-$ are the means of the two samples in $B_S^+$ and $B_S^-$, respectively. The value $v^*$ is a natural choice for constructing a loss function that aims to distinguish the boundary regions. However, owing to the different compositions of the target organ and background tissues, the variance of the grayscale values in the $B_S^+$ and $B_S^-$ regions may differ significantly. Consequently, a simple comparison of the means of the two regions is inadequate. We propose a statistical approach based on the two-sample t-test to address this issue \citep{ruxton2006unequal}.


In particular, the grayscale values of $B^+_S$ and $B^-_S$ can be considered as two samples. Recall that the mean values of the two samples are $\bar{v}_+$ and $\bar{v}_-$, respectively. Furthermore, we denote the standard deviations of the two samples as $s_+$ and $s_-$, respectively. Subsequently, the t-test statistic can be defined as
\begin{equation}
\label{Eq:t-statitic}
    t^*=(\bar{v}_+-\bar{v}_-)\Big/\sqrt{\dfrac{s_+^2}{n_+}+\dfrac{s_-^2}{n_-}},
\end{equation}
where $n_+$ and $n_-$ are the sizes of the respective samples. Compared with \eqref{eq:spta-idea}, the newly defined $t^*$ considers the variance term, and thus, the influence of heteroscedasticity, i. e. difference in variance, is accounted for when comparing the two regions. In statistics, the value $t^*$ is often used in hypothesis testing to explore whether two samples share equal mean values. In this study, we use $t^*$ to characterize the difference in the grayscale value distributions inside and outside the boundary of the segmented region. With a larger $|t^*|$ value, $\partial S$ is more likely to be the true boundary. Therefore, the value of $|t^*|$ is a good indicator for boundary detection. We use $1/|t^*|$ as the loss function for consistency with the model optimization direction.

\subsection{PTA Loss}
\label{section:PTA}

\begin{figure}[h]
    \centering
    \includegraphics[scale=0.6]{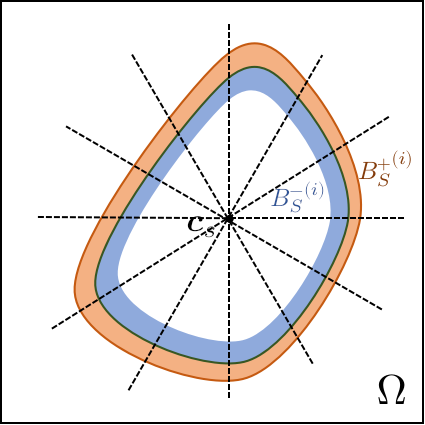}
    \caption{Illustration of dividing entire region into several parts.}
    \label{fig:piecewise_B+B-}
\end{figure}

As most segmentation targets (e.g., organs) and their surrounding backgrounds are asymmetrical, attention must be paid to the changing patterns along the boundaries.
We divide the entire boundary into several parts to adapt to such patterns. Specifically, let $\boldsymbol{c}_S$ be the center of region $S$, which can be calculated as $\boldsymbol{c}_S=\sum_{\boldsymbol{x}\in S}{\boldsymbol{x}}/{|S|}$. Assume that $\partial S$ can be divided into $K$ parts, where $K$ is predefined. Accordingly, $B^+_S$ and $B^-_S$ can be divided into a sequence of subregions $\{({B^+_S}^{(i)}, {B^-_S}^{(i)})\}_{i=1}^K$, as shown in Figure \ref{fig:piecewise_B+B-}.

Mathematically, $\{({B^+_S}^{(i)}, {B^-_S}^{(i)})\}$ can be defined as follows:
\begin{equation}
\begin{aligned}
{B^+_S}^{(i)}=\{\boldsymbol{x}\in B^+_S:\langle\boldsymbol{x},\boldsymbol{c}_S\rangle\in (\frac{2(i-1)\pi}{K},\frac{2i\pi}{K}]\},\\
{B^-_S}^{(i)}=\{\boldsymbol{x}\in B^-_S:\langle\boldsymbol{x},\boldsymbol{c}_S\rangle\in (\frac{2(i-1)\pi}{K},\frac{2i\pi}{K}]\}.
\end{aligned}
\end{equation}
For each pair of $({B^+_S}^{(i)}, {B^-_S}^{(i)})$, a two-sample t-test statistic can be obtained, similar to (\ref{Eq:t-statitic}). Let $t^{(i)}=t^*({B^+_S}^{(i)}, {B^-_S}^{(i)})$ be the corresponding t-test statistic on the region $({B^+_S}^{(i)}, {B^-_S}^{(i)})$. Subsequently, a sequence of statistics can easily be obtained as $\boldsymbol{t}=(t^{(1)}, t^{(2)},\cdots, t^{(K)})$. Then, we define $\mathcal{L}_{PT}=K^{-1}\sum_{i=1}^K\mathcal{L}_{PT}^{(i)}$, where $\mathcal{L}_{PT}^{(i)}=1/t^{(i)}$. Here, $\mathcal{PT}$ represents the \emph{piecewise t-test}, which can measure whether the grayscale value distributions of the pixels around the boundary are significantly different.

Practically, the loss $\mathcal{L}_{PT}$ can be combined with any classical loss function. The classical loss functions are useful for detecting the general location of the target region, whereas the new loss $\mathcal{L}_{PT}$ is helpful for accurately detecting the boundary. Let $\mathscr{L}=\{\mathcal{L}(G,S): G,S\subset \Omega\}$ define the class of loss functions. Subsequently, we combine our proposed $\mathcal{L}_{PT}$ with any loss function $\mathcal{L}\in\mathscr{L}$. This leads to the following \emph{PTA loss}:
\begin{equation}
\label{Eq:PTA}
    \mathcal{L}_{PTA}=\mathcal{L}+\lambda \mathcal{L}_{PT} ,\quad \mathcal{L}\in\mathscr{L},
\end{equation}
where $\lambda$ denotes a hyperparameter that balances the two loss functions. In other words, the hyperparameter $\lambda$ can balance the segmentation accuracy both in the interior region and at the boundary, so that the model can be updated to achieve better boundary accuracy with acceptable compromise in the interior and exterior predictions.
Particularly, in the case that the variances of $I({B_S^+}^{(i)})$ and $I({B_S^-}^{(i)})$ are observed to be similar. Then our $\mathcal{L}_{PT}$ term
becomes equivalent to the  difference value $v^{(i)}=|\bar{v}^{(i)}_+-\bar{v}^{(i)}_-|$ for each piece, which we name as the simplified PTA loss (sPTA).



%
%

\section{EXPERIMENT ON SYNTHETIC DATA}

\begin{figure}[H]
    \centering
     \begin{subfigure}{.281\textwidth}
        \centering
        \includegraphics[width=\textwidth]{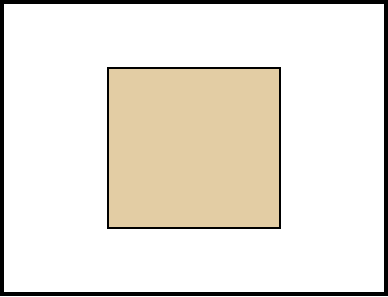}
        \caption{{\sc Case 1}: correct}
        \label{fig:case-correct}
    \end{subfigure}\quad
    \begin{subfigure}{.281\textwidth}
        \centering
        \includegraphics[width=\textwidth]{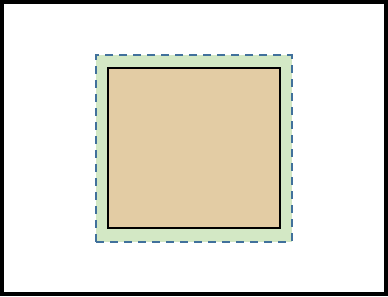}
        \caption{{\sc Case 2}: large}
        \label{fig:case-larger}
    \end{subfigure}\quad
    \begin{subfigure}{.281\textwidth}
        \centering
        \includegraphics[width=\textwidth]{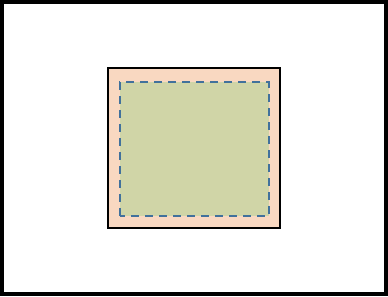}
        \caption{{\sc Case 3}: small}
        \label{fig:case-small}
    \end{subfigure}

    \begin{subfigure}{.281\textwidth}
        \centering
        \includegraphics[width=\textwidth]{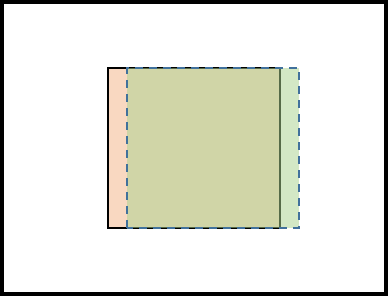}
        \caption{{\sc Case 4}: horizontal}
        \label{fig:case-hbias}
    \end{subfigure}\quad
    \begin{subfigure}{.281\textwidth}
        \centering
        \includegraphics[width=\textwidth]{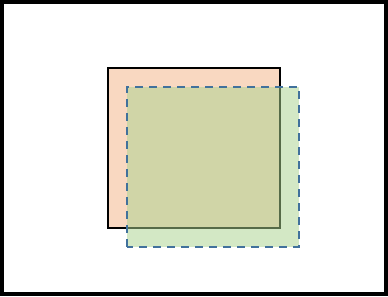}
        \caption{{\sc Case 5}: diagonal}
        \label{fig:case-hvbias}
    \end{subfigure}
    \caption{Five simulation cases. In each case, the ground-truth region $G$ and segmentation region $S$ are colored in orange and green, respectively.}
    \label{fig:cases}
\end{figure}

We simulated different scenarios to determine the influence of $\mathcal{L}_{PTA}$ on the model optimization, and thus, validate its effectiveness for medical image segmentation. We assumed that the spatial region was $\Omega=[0,200]^2$. A simple geometry ($60\times60$-pixel square) was used to mimic the target organ, which was located at $G=[70,130]^2$. We let the grayscale pixels inside and outside the target organ be generated from two different normal distributions $N(3.5,2)$ and $N(0,2)$, respectively, to generate the fake grayscale map. We considered five cases, each of which had a small offset around the target organ. Figure \ref{fig:cases} depicts the five cases. Specifically, in {\sc Case 1}, the segmentation region was exactly the same as the organ; that is, $S_1=[70, 130]^2$. In {\sc Case 2} and {\sc Case 3}, the segmentation region was larger or smaller than the organ, and we defined $S_2=[68, 132]^2$ and $S_3=[72,128]^2$. In {\sc Case 4} and {\sc Case 5}, the segmentation region moved horizontally or diagonally from the organ. To reflect these two scenarios, we defined the segmentation regions as $S_4=[75,135]\times[70,130]$ and $S_5=[75,135]^2$. The final four cases simulated boundary blurring, which is one major restriction of the model performance.

The entire boundary region was divided into $K=10$ pieces. The values of $\mathcal{L}_{PT}^{(i)}$ and averaged $\mathcal{L}_{PT}$ were computed for all $K=10$ pieces and are shown in Table \ref{tab:simulation-case}. The $F_1$ score is also reported as a measure of the segmentation performance. As indicated in Table \ref{tab:simulation-case}, $L_{ST}$ was the smallest when the segmentation was accurate. However, $\mathcal{L}_{PT}$ increased in {\sc Case 2} and {\sc Case 3}. Furthermore, the horizontal and diagonal moves led to a sharp increase in $\mathcal{L}_{PT}$. These results suggest that the loss $L_{ST}$ can guide the model parameter optimization in the correct direction. An analysis of the 10-part $\mathcal{L}_{PT}^{(i)}$ values reveals that, in the case in which similarities and central symmetry existed among the 10 parts, the value of $\mathcal{L}_{PT}^{(i)}$ for each part became relatively similar. In the case in which the different parts faced varying differences in the gray distributions, the part with the larger $\mathcal{L}_{PT}^{(i)}$ was the interlaced part of the offset. The part with a large $\mathcal{L}_{PT}^{(i)}$ could reflect the characteristics of the offset to a certain extent, which indicates that the $\mathcal{L}_{PT}$ loss also has a geometric property.

\begin{table}[h]
    \centering
    \footnotesize
    \caption{Values of $\mathcal{L}_{PT}^{(i)}$ and $\mathcal{L}_{PT}$ in five simulation cases. The $F_1$ score is also reported as a measure of the segmentation performance.}
    \begin{tabular}{lccccc}
    \hline
        & {\sc Case 1} & {\sc Case 2} & {\sc Case 3} & {\sc Case 4} & {\sc Case 5}\\
        \hline
        $\mathcal{L}_{PT}^{(1)}$ & 0.278 & 0.572 & 0.371 & 0.277 & 2.011 \\
        $\mathcal{L}_{PT}^{(2)}$ & 0.303 & 0.464 & 0.465 & 2.417 & 3.590 \\
        $\mathcal{L}_{PT}^{(3)}$ & 0.328 & 0.451 & 0.464 & 5.453 & 4.483 \\
        $\mathcal{L}_{PT}^{(4)}$ & 0.296 & 0.526 & 0.411 & 1.633 & 1.804 \\
        $\mathcal{L}_{PT}^{(5)}$ & 0.280 & 0.585 & 0.442 & 0.311 & 2.654 \\
        $\mathcal{L}_{PT}^{(6)}$ & 0.277 & 0.447 & 0.495 & 0.276 & 2.243 \\
        $\mathcal{L}_{PT}^{(7)}$ & 0.278 & 0.460 & 0.431 & 0.678 & 6.067 \\
        $\mathcal{L}_{PT}^{(8)}$ & 0.255 & 0.582 & 0.391 & 3.351 & 2.167 \\
        $\mathcal{L}_{PT}^{(9)}$ & 0.260 & 0.472 & 0.437 & 1.103 & 1.945 \\
        $\mathcal{L}_{PT}^{(10)}$ & 0.287 & 0.553 & 0.431 & 0.325 & 2.136 \\ \hline
        $\mathcal{L}_{PT}$ & 0.285 & 0.512 & 0.434 & 1.583 & 2.911 \\
        $F_1$ score & 1.000 & 0.935 & 0.931 & 0.916 & 0.840 \\\hline
    \end{tabular}
    \label{tab:simulation-case}
\end{table}

\begin{figure}[h]
    \centering
    \includegraphics[width=\textwidth]{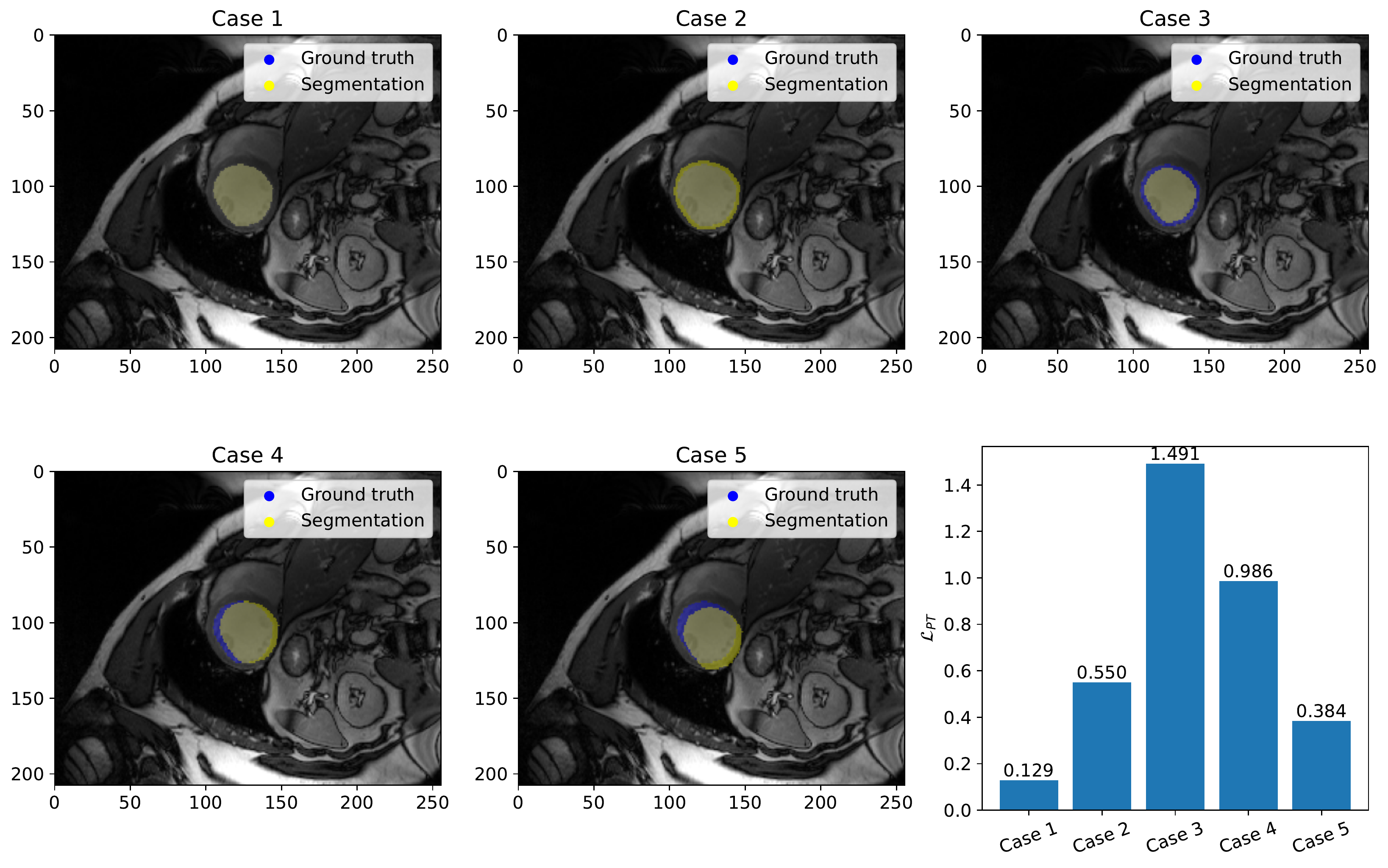}
    \caption{Five simulation cases on real datasets. In each case, the ground-truth region $G$ and segmentation region $S$ are colored in blue and yellow, respectively. A bar plot is used to illustrate the results of $\mathcal{L}_{PT}$.}
    \label{fig:simulation-real}
\end{figure}

\clearpage

To demonstrate the operational direction of the $\mathcal{L}_{PT}$ loss further, we recreated the five cases above on a real medical image, as discussed in Section 4. Specifically, we randomly selected one medical image from a real cardiac cine MRI dataset. Following the procedure used above, we created five incorrect segmentation cases by setting the offset to 3 pixels. We computed the $\mathcal{L}_{PT}$ loss for each case. Figure \ref{fig:simulation-real} depicts the five cases and the resulting $\mathcal{L}_{PT}$. Although the incorrect segmentation region was generally consistent with the ground-truth region for most pixels, its segmentation error on the boundary made the resulting $\mathcal{L}_{PT}$ positive. This demonstrates the sensitivity of the proposed loss function to the boundary.

\section{EXPERIMENT ON REAL DATA}

\subsection{Data Description}

\begin{figure}[h]
    \centering
    \includegraphics[width=0.3\textwidth]{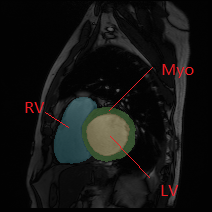}
    \caption{Illustration of segmentation areas and labels. The blue, yellow, and green areas represent the LV, RV, and Myo, respectively.}
    \label{fig:segmentation_sample}
\end{figure}

We conducted experiment on public cardiac cine MRI training data from the ACDC Challenge (\url{https://www.creatis.insa-lyon.fr/Challenge/acdc}). This dataset includes cardiac MRI images of 100 patients, with segmentation masks of the left ventricle (LV), right ventricle (RV), and myocardium (Myo) at the end-diastolic and end-systolic phases for each patient. Figure \ref{fig:segmentation_sample} presents the three segmentation areas of the organ. In the preprocessing stage, we resampled all images to a common in-plane resolution of 1.37 mm $\times$ 1.37 mm. However, we did not perform any resampling in the through-plane direction to avoid a loss of information in the upsampling and downsampling steps. Subsequently, following the practice of \cite{CardiacMRI}, we cut all resampled images into constant-sized images, i.e., 212 $\times$ 212 for the U-Net model and 224 $\times$ 224 for the other models, padding zeros if necessary to meet the image requirements. Last we divided the dataset into two parts: 80 patients as the training set and 20 patients as the test set.

\subsection{Models and Loss Functions}

We adopted three representative network architectures for the image segmentation task to compare and prove the effectiveness of PTA loss. These are briefly introduced as follows:

\begin{itemize}
\item\textbf{Fully convolutional network (FCN)}
\citep{Long_2015_CVPR}. This is the first end-to-end CNN network for pixel-level prediction, which uses a deconvolutional layer to upsample the feature map of the final convolutional layer. In this manner, a prediction can be generated for each pixel while preserving the spatial information in the original input image. Finally, it performs pixel-by-pixel classification using softmax on the upsampled feature map. The FCN model used in this work was based on the VGG-16 architecture and used three skip connections \citep{CardiacMRI}.

\item\textbf{U-Net}
\citep{Ronneberger2015}. This model was developed from the FCN. With the aid of concatenation as a fusion of feature maps on the same layer, the U-shaped encoder--decoder structure of U-Net means that it can integrate the contextual and located information. In this work, U-Net was modified slightly following the instructions of Baumgartner et al.\cite{CardiacMRI}. Furthermore, it was adapted with the number of feature maps in the transpose convolutions of the upsampling path, which was set as the number of classes.

\item\textbf{nnUnet}
\citep{isensee2018nnu}. This network is an adaptive framework based on U-Net. The network architecture is similar to the traditional U-Net, but specifically optimized in the preprocessing, training, inference, and post-processing steps to achieve better results. Data enhancement methods such as random rotation, random scaling, random elastic transformation, gamma correction, and mirroring were used to construct the adaptive network.

\end{itemize}

We used six different loss functions based on the three network architectures described above: the standard pixel-level CE, WCE, DSC, active contour loss (ACL) \citep{Chen_2019_CVPR,Le2021OffsetCurve}, PTA loss, and its simplified version, sPTA. The CE, WCE, and DSC are traditional loss functions, whose definitions are presented in Section 2. As mentioned in Section 1, the ACL loss function is additive and combines the traditional loss functions with a rewritten energy function of the active contour model \citep{Chen_2019_CVPR,Le2021OffsetCurve}. Our proposed PTA is also an additive loss function, and its combination with the traditional loss functions can lead to a smoother boundary and improved performance compared to the original loss. We compared the performance of our proposed PTA with that of earlier work on the ACL.

Each model was trained using an Adam optimizer. The learning rate of each parameter was adjusted dynamically by adapting the first- and second-order moment estimations of the gradients. The batch size was set to 10, and we used early stopping to avoid overfitting; that is, the experiment was terminated if no better results emerged after 15 consecutive epochs \citep{CardiacMRI}. The hyperparameter values were selected using a grid search. Following common practice, the weights in the WCE loss were set to 0.3 for the three foreground classes and 0.1 for the background region. The value of $\lambda$ was set to 10 for the ACL loss and 3 for our proposed sPTA and PTA losses. Moreover, we set $K=10$ to divide the $B^+$ and $B^-$ regions.

\subsection{Evaluation Metrics}

We evaluated the segmentation performance using the following metrics: the DSC, precision and recall, average symmetric surface distance (ASSD), and Hausdorff distance (HD). These metrics are commonly used for segmentation evaluations \citep{EvaluationMethod_1997, Chen_2019_CVPR}. Specifically, the DSC measures the overlapping region between the segmentation result and ground truth, whereas the HD and ASSD measure the distance between the two boundaries of the target region in the segmentation result and that in the ground truth. We provide a detailed description of the metrics in the following.

\begin{itemize}
\item\textbf{DSC.} The DSC is defined as $DSC = 1-2|G\cap S|/(|G|+|S|).$ It measures the proportion of the intersection over the union of the target region $G$ and segmentation region $S$. It is equal to the $F_1$ score.

\item\textbf{Precision \& recall.} The precision is defined as $|G\cap S|/(|S|).$ It measures the proportion of the true positive prediction over the segmentation region $S$. The recall is defined as $|G\cap S|/(|G|).$ It measures the proportion of the true positive prediction over the target region $G$.

\item\textbf{ASSD.}
The ASSD is defined as $\mathcal ASSD= \{\sum_{\boldsymbol{x}\in{\partial S}} d(x,\partial G)+\sum_{\boldsymbol{x}\in{\partial G}} d(x,\partial S)\}/(\int_{\partial G}dG+\int_{\partial S}dS).$ It is also known as the distance of the mean surface. It calculates the average value of the distances from every point in one of the boundary point sets of the ground truth and segmentation result to the other point set.

\item\textbf{HD.}
The HD is defined as $\mathcal HD= \max\{\sup_{\boldsymbol{x}\in{\partial S}} d(x,\partial G),\sup_{\boldsymbol{x}\in{\partial G}} d(x,\partial S)\}.$ As opposed to the ASSD, the HD calculates the supremum of the distances from each point pair.
\end{itemize}

\subsection{Experimental Results}

\begin{table}[p]
    \centering
    \footnotesize
\caption{Experimental results of different loss functions under FCN, U-Net, and nnU-Net models. Three base loss functions (CE, WCE, and DSC) were used, each of which was combined with the ACL, PTA and sPTA loss functions. The segmentation performance was evaluated using the DSC, precision, recall, HD, and ASSD metrics under each loss function.}
\label{table:Result1}
    \begin{tabular}{cccccccc} 
    \hline
    \hline
        Network & Base loss & Additive loss & DSC & Precision & Recall & HD & ASSD  \\
        \hline
        \multirow{12}{*}{FCN}&\multirow{4}{*}{CE} & $-$ & 0.9058  & 0.9058  & 0.9101  & 9.6963  & 0.6858   \\
        && ACL & 0.8948  & 0.8981  & 0.8969  & 10.1721  & 0.7487   \\
        && sPTA & 0.9135  & \textbf{0.9117}  & 0.9195  & 9.3281  & \textbf{0.5885}   \\
        && PTA & \textbf{0.9139}  & 0.9099  & \textbf{0.9217}  & \textbf{9.1173}  & 0.5915   \\
        \cline{2-8}
        &\multirow{4}{*}{WCE} & $-$ & 0.9077  & 0.8955  & 0.9259  & 9.6589  & 0.7126   \\
        && ACL & 0.8744  & 0.8651  & 0.8902  & 11.2245  & 0.8775  \\
        && sPTA & 0.9073  & 0.8878  & \textbf{0.9340}  & \textbf{9.4115}  & 0.7196   \\
        && PTA & \textbf{0.9125}  & \textbf{0.9041}  & 0.9260  & 9.4953  & \textbf{0.6690}  \\
        \cline{2-8}
        &\multirow{4}{*}{DSC} & $-$ & 0.8897  & 0.8651  & \textbf{0.9287}  & 10.8005  & 1.0457   \\
        && ACL & 0.8914  & 0.8810  & 0.9102  & \textbf{10.2331}  & 0.9793   \\
        && sPTA & 0.8968  & 0.8877  & 0.9156  & 10.5536  & \textbf{0.8950}   \\
        && PTA & \textbf{0.8975}  & \textbf{0.8878}  & 0.9170  & 10.5830  & 0.9101   \\
    \hline
        \multirow{12}{*}{Unet}&\multirow{4}{*}{CE} & $-$ & 0.9036  & 0.9015  & 0.9111  & 10.1558  & 0.7901   \\
        && ACL & 0.9072  & 0.9031  & 0.9175  & 10.6839  & 0.7217  \\
        && sPTA & 0.9131  & 0.9077  & \textbf{0.9236}  & 9.9370  & 0.6783  \\
        && PTA & \textbf{0.9144}  & \textbf{0.9192}  & 0.9142  & \textbf{9.8881}  & \textbf{0.6469}  \\
        \cline{2-8}
        &\multirow{4}{*}{WCE} & $-$ & 0.9088  & 0.8958  & 0.9282  & 9.5619  & 0.6975   \\
        && ACL & 0.9034  & 0.8892  & 0.9249  & 10.5752  & 0.7561   \\
        && sPTA & 0.9098  & 0.9003  & 0.9250  & \textbf{9.3703}  & 0.6870    \\
        && PTA & \textbf{0.9136}  & \textbf{0.9021}  & \textbf{0.9295}  & 9.6260  & \textbf{0.6461}   \\
        \cline{2-8}
        &\multirow{4}{*}{DSC} & $-$ & 0.8831  & 0.8621  & 0.9222  & 12.2202  & 1.1885  \\
        && ACL & 0.8902  & 0.8666  & \textbf{0.9301}  & 11.1434  & 1.1238   \\
        && sPTA & 0.8918  & 0.8715  & 0.9259  & 10.7886  & 1.0775   \\
        && PTA & \textbf{0.9124}  & \textbf{0.9019}  & 0.9277  & \textbf{9.4757}  & \textbf{0.6496}   \\
    \hline
        \multirow{12}{*}{nnUnet}&\multirow{4}{*}{CE} & $-$ & 0.9271  & 0.9133  & 0.9098  & 10.8643  & 0.1813   \\
        && ACL & 0.9296  & 0.9162  & 0.9148  & 9.9651  & 0.1741  \\
        && sPTA & 0.9285  & 0.9130  & 0.9115  & 10.9678  & 0.3291   \\
        && PTA & \textbf{0.9311}  & \textbf{0.9165}  & \textbf{0.9151}  & \textbf{9.8666}  & \textbf{0.1699}  \\
        \cline{2-8}
        &\multirow{4}{*}{WCE} & $-$ & 0.9282  & 0.9004  & 0.9349  & 10.5241  & 0.1508   \\
        && ACL & 0.9291  & \textbf{0.9098}  & 0.9297  & \textbf{9.1217}  & \textbf{0.1442}   \\
        && sPTA & 0.9285  & 0.8985  & 0.9336  & 10.4629  & 0.1563   \\
        && PTA & \textbf{0.9312}  & 0.9019  & \textbf{0.9350}  & 10.5612  & 0.1486   \\
        \cline{2-8}
        &\multirow{4}{*}{DSC} & $-$ & 0.9280  & 0.9196  & \textbf{0.9277}  & 13.5788  & 0.2859   \\
        && ACL & 0.9308  & 0.9187  & 0.9192  & 13.1956  & \textbf{0.1904}   \\
        && sPTA & 0.9312  & 0.9193  & 0.9270  & 12.4862  & 0.3210   \\
        && PTA & \textbf{0.9334}  & \textbf{0.9217}  & 0.9265  & \textbf{11.7988}  & 0.2470   \\
    \hline
    \hline
    \end{tabular}
\end{table}
Table \ref{table:Result1} summarizes the detailed experimental results of the different loss functions under the FCN, U-Net, and nnU-Net. We considered the three base loss functions (the CE, WCE, and DSC), each of which was combined with the ACL, PTA, and sPTA loss functions. Under each loss function, the image segmentation performance was evaluated using the DSC, precision, recall, HD, and ASSD metrics.
The precision and recall  are defined according to the macro-average, and each of them alone should not be interpreted as performance measure individually. The following conclusions can be drawn from the results: First, under different base loss functions and model architectures, the proposed PTA or sPTA always improved the base loss for nearly all evaluation metrics. This improvement was more obvious for the FCN and U-Net models, the architectures of which are simpler than that of nnU-Net.
The PTA and sPTA losses exhibited superior performance in terms of the the DSC than the other loss functions.
In particular, the PTA and sPTA outperformed the ACL loss in most cases. Note that the ACL is also an additive loss function and focuses on the boundary region. This finding indicates that the difference inside and outside the boundary can be better captured using the two-sample t-test method compared to differentiation operators. Finally, the PTA outperformed the sPTA in most cases. This result implies that, although the sPTA is simpler to compute, the PTA should be the best option for most cases.

\begin{figure}[h]
    \centering
    \includegraphics[width=.8\textwidth]{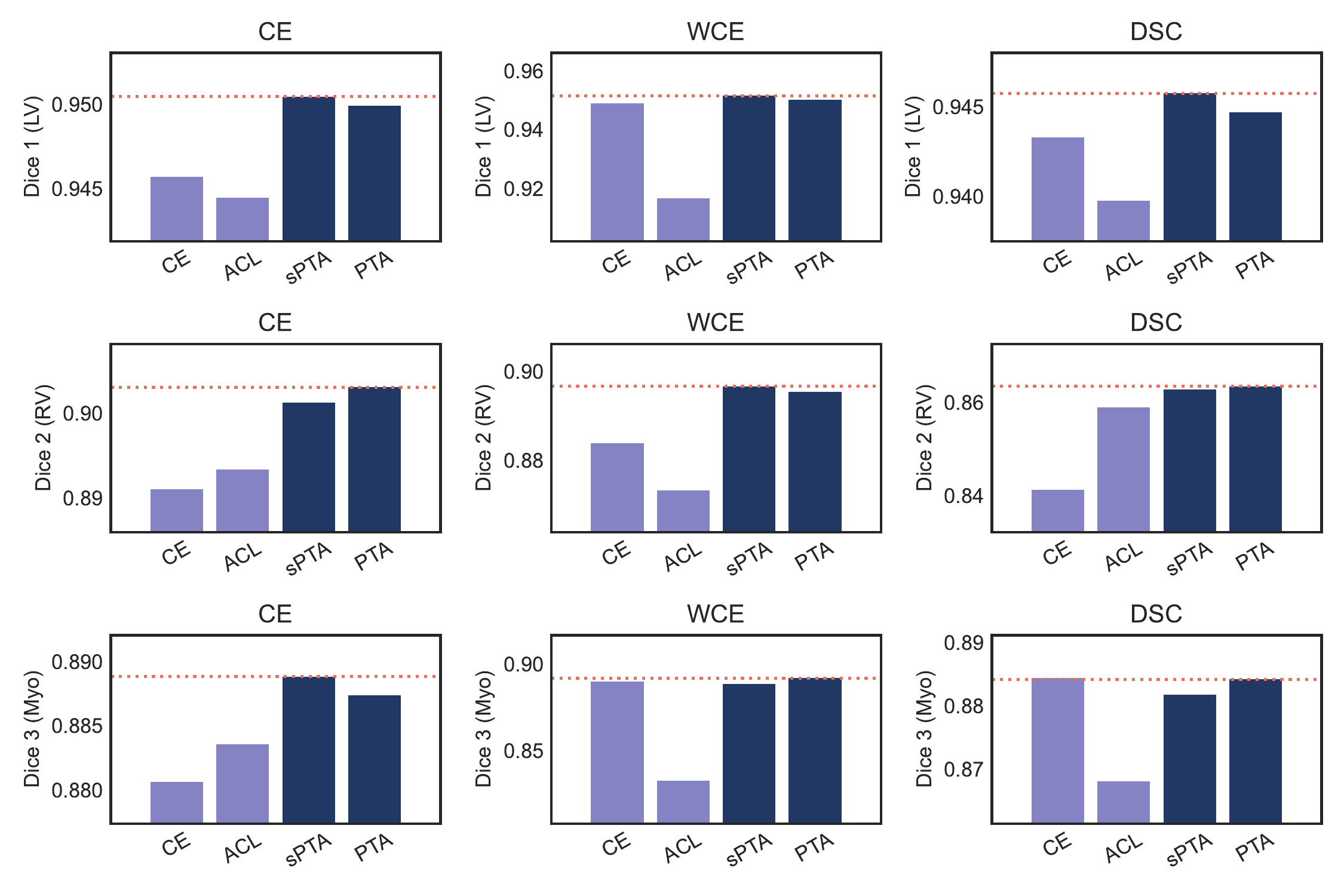}
    \caption{DSC results of three classes: LV, RV, and Myo, under FCN model.}
    \label{fig:comp-fcn}
\end{figure}

\begin{figure}[h]
    \centering
    \includegraphics[width=.8\textwidth]{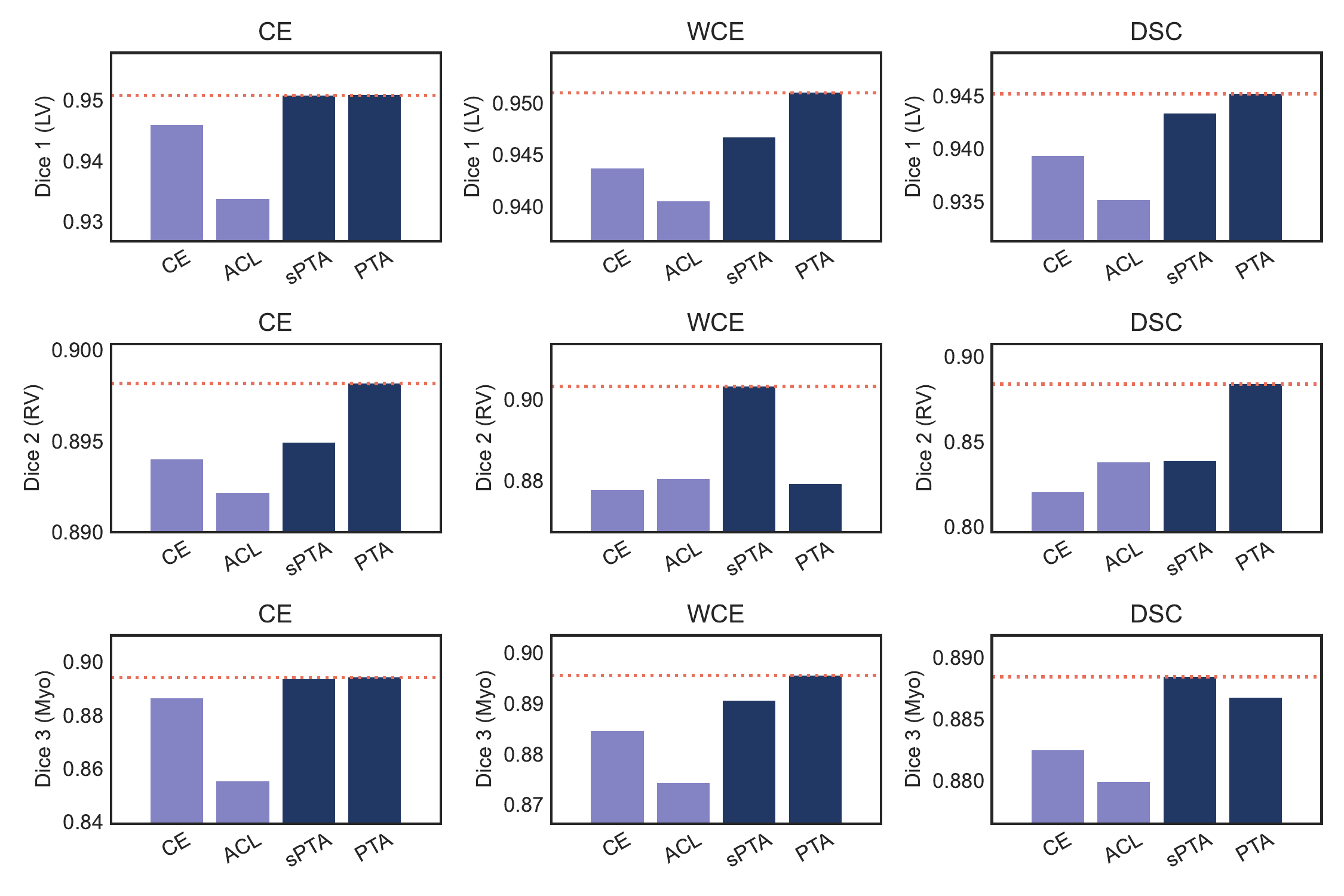}
    \caption{DSC results of three classes: LV, RV, and Myo, under U-Net model.}
    \label{fig:comp-unet}
\end{figure}

\begin{figure}[hpt]
    \centering
    \includegraphics[width=.8\textwidth]{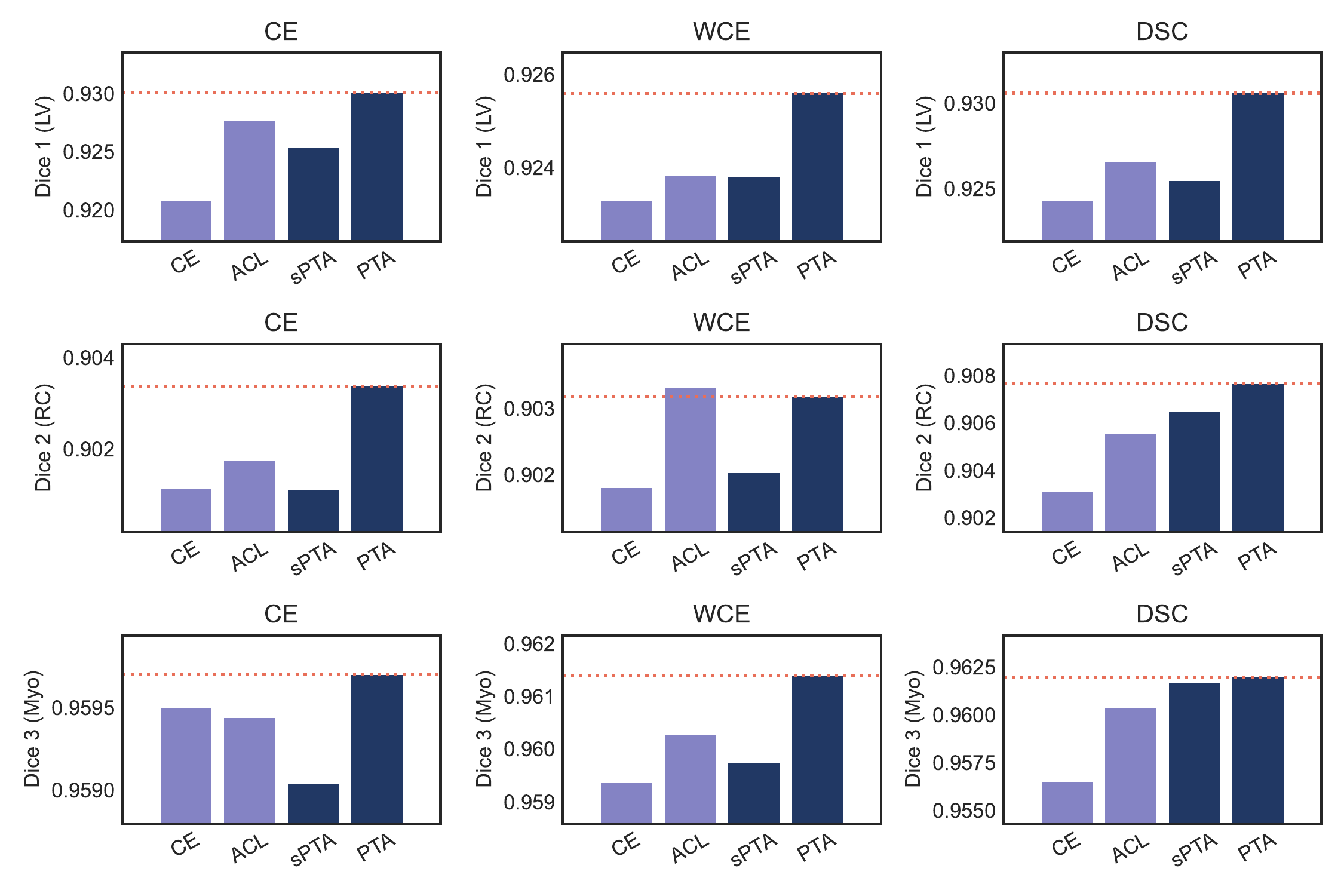}
    \caption{DSC results of three classes: LV, RV, and Myo, under nnU-Net model.}
    \label{fig:comp-nnunet}
\end{figure}

Next, we focus on the segmentation performance of the three classes (LV, RV, and Myo). Taking the DSC as an example, Figures \ref{fig:comp-fcn} to \ref{fig:comp-nnunet} present the DSC results for each class under the different loss functions and model architectures. It is obvious that the segmentation accuracy could be improved by using the PTA or sPTA loss functions in most cases. This improvement was much more obvious when the CE was used as the base loss. This is probably because the CE loss depends less on the geometric information of the images. Thus, the incorporation of the boundary information using the two-sample t-test could provide additional insight. A comparison of the different classes reveals that the RV class exhibited the greatest improvement when using PTA or sPTA losses. The RV region had the longest boundary among the three classes. Therefore, its segmentation performance was the most beneficial from the perspective of accurately distinguishing the boundary.

\begin{figure*}[h]
    \begin{subfigure}[t]{0.24\textwidth}
           \centering
           \includegraphics[width=\textwidth]{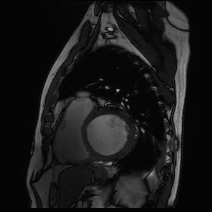}
            \caption{Raw image}
            \label{fig:raw1}
    \end{subfigure}
    \begin{subfigure}[t]{0.24\textwidth}
            \centering
            \includegraphics[width=\textwidth]{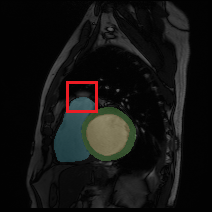}
            \caption{Ground truth}
            \label{fig:groundtruth1}
    \end{subfigure}
    \begin{subfigure}[t]{0.24\textwidth}
            \centering
            \includegraphics[width=\textwidth]{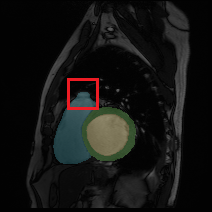}
            \caption{WCE result}
            \label{fig:base1}
    \end{subfigure}
	\begin{subfigure}[t]{0.24\textwidth}
            \centering
            \includegraphics[width=\textwidth]{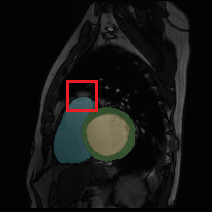}
            \caption{PTA result}
            \label{fig:our1}
    \end{subfigure}

    \begin{subfigure}[t]{0.24\textwidth}
           \centering
           \includegraphics[width=\textwidth]{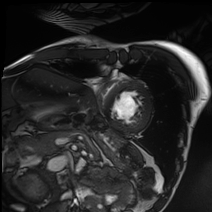}
            \caption{Raw image}
            \label{fig:raw2}
    \end{subfigure}
    \begin{subfigure}[t]{0.24\textwidth}
            \centering
            \includegraphics[width=\textwidth]{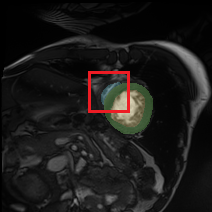}
            \caption{Ground truth}
            \label{fig:groundtruth2}
    \end{subfigure}
    \begin{subfigure}[t]{0.24\textwidth}
            \centering
            \includegraphics[width=\textwidth]{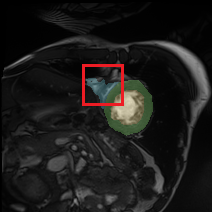}
            \caption{WCE result}
            \label{fig:base2}
    \end{subfigure}
	\begin{subfigure}[t]{0.24\textwidth}
            \centering
            \includegraphics[width=\textwidth]{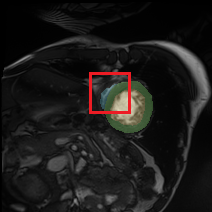}
            \caption{PTA result}
            \label{fig:our2}
    \end{subfigure}
    \caption{Illustration of two representative examples. The raw image, ground truth segmentation, and segmentation results using the WCE loss and our PTA loss are presented. The three classes (LV, RV, and Myo) are indicated in different colors.}
	\label{fig:realcase}
\end{figure*}

\newpage
Finally, we selected two representative examples to illustrate the segmentation performance of the PTA loss. Figure \ref{fig:realcase} presents the raw image, ground-truth segmentation, and segmentation results using the WCE loss and our PTA loss under the U-Net model architecture for each example. Compared to the segmentation results using the WCE loss, our proposed PTA loss resulted in a smoother segmentation boundary that was closer to the ground truth. This finding again demonstrates that the PTA loss can achieve more distinct and accurate boundaries for the target region.

\newpage
\section{Conclusion}
We have proposed the PTA loss for medical image segmentation. With a focus on boundary segmentation, the PTA loss applies the statistical hypothesis testing method to measure the difference between the gray-value distributions inside and outside the boundary region. The PTA loss is more effective in detecting tiny differences, and thus, results in a smoother and more accurate segmentation boundary compared to previous loss functions. Our experiments on synthetic data demonstrated the direction of action of the PTA loss. Furthermore, experiments on a public cardiac cine MRI dataset revealed the promising segmentation performance of the PTA loss. Our work can be extended in several directions. First, the proposed PTA loss can easily be extended to color images using the three RGB channels. A possible solution is to compute the PTA loss for each channel and then sum them. More complex multivariate hypothesis-testing statistics can also be considered. Second, as the proposed PTA loss function is an additive loss function, it can be combined with more traditional loss functions and used for additional deep learning models.

\section*{Disclosures}

The authors declare no conflicts of interest.


\bibliography{sample}






\end{document}